\documentclass[twocolumn]{webofc}

\usepackage[varg]{txfonts}   
\usepackage{url}
\usepackage{physics, mathtools, braket}
\usepackage{soul}

\usepackage{hyperref}
\hypersetup{
    colorlinks = true,
    citecolor  = blue,
    linkcolor  = blue,
    urlcolor  = blue
}

\begin{document}

\newcommand{\Sch}{ Schr\"{o}dinger }
\newcommand{\ie}{\textit{i.e.}}
\newcommand{\eg}{\textit{e.g.}}

\newcommand{\FM}[1]{{\color{magenta} #1}}
\newcommand{\can}[1]{{\color{red}\sout{#1}}}
\newcommand{\FB}[1]{{\color{orange} #1}}
\newcommand{\AT}[1]{{\color{blue}#1}}

\title{Recent advances in coupled cluster computations of open-shell atomic nuclei}

\author{
    \firstname{Francesco} \lastname{Marino} \inst{1}
    \fnsep \thanks{
    \email{frmarino@uni-mainz.de}
    } 
\and
    \firstname{Francesca} \lastname{Bonaiti}\inst{1,2,3}
\and
    \firstname{Sonia} \lastname{Bacca}\inst{1,4}
\and
    \firstname{Pepijn} \lastname{Demol}\inst{5}
\and
    \firstname{Thomas} \lastname{Duguet}\inst{6,5}
\and
    \firstname{Gaute} \lastname{Hagen}\inst{3,7}
\and
    \firstname{Gustav} \lastname{Jansen}\inst{8,3}
\and
    \firstname{Alexander} \lastname{Tichai}\inst{9,10,11}
}

\institute{
    Institut f\"{u}r Kernphysik and PRISMA+ Cluster of Excellence, Johannes Gutenberg-Universit\"{a}t Mainz, 55099 Mainz, Germany
\and 
    Facility for Rare Isotope Beams, Michigan State University, East Lansing, MI 48824, USA
\and
    Physics Division, Oak Ridge National Laboratory, Oak Ridge, TN 37831, USA
\and
    Helmholtz-Institut Mainz, Johannes Gutenberg-Universität Mainz, D-55099 Mainz, Germany
\and
    KU Leuven, Instituut voor Kern- en Stralingsfysica, 3001 Leuven, Belgium
\and
    IRFU, CEA, Université Paris-Saclay, 91191 Gif-sur-Yvette, France
\and
    Department of Physics and Astronomy, University of Tennessee, Knoxville, TN 37996, USA
\and
    National Center for Computational Sciences, Oak Ridge National Laboratory, Oak Ridge, TN 37831, USA
\and
    Technische Universität Darmstadt, Department of Physics, 64289 Darmstadt, Germany
\and
    ExtreMe Matter Institute EMMI, GSI Helmholtzzentrum für Schwerionenforschung GmbH, 64291 Darmstadt, Germany
\and
    Max-Planck-Institut für Kernphysik, Saupfercheckweg 1, 69117 Heidelberg, Germany
}

\abstract{
In this contribution, we report on recent progress in coupled-cluster simulations of open-shell atomic nuclei using interactions consistently derived from chiral effective field theory. In particular, we compare different coupled-cluster approaches by computing binding energies and electric dipole polarizabilities in medium-mass calcium isotopes. }

\maketitle

\section{Introduction}
\label{intro}
Coupled-cluster (CC) theory~\cite{Hagen2014Review,ShavittBartlett} is a powerful many-body approach for studying the structure and dynamics of atomic nuclei. Its mild computational scaling enables the extension to heavy systems (see e.g., Refs.~\cite{Gysbers2019,Hu2022}). Through the computation of the response functions, it further allows to study electroweak properties~\cite{Bacca2013}.
In this work, we document recent progress in applying this method to open-shell nuclei. In particular, we report calculations of binding energies and electric dipole polarizabilities in the Ca isotopic chain, with a particular focus on benchmarking various CC formulations.

\section{Review of the theory}
\label{sec: theory}
The single-reference coupled-cluster framework~\cite{Hagen2014Review,ShavittBartlett} employs an exponential parametrization of the many-body ground state,
\begin{align}
    \label{eq: cc gs ansatz}
    \ket{\Psi_0} = e^{ \hat{T} } \ket{ \Phi_0 },
\end{align}
where $\ket{ \Phi_0 }$ is a reference state, commonly obtained from a variational mean-field solution, \eg{}, a Hartree-Fock Slater determinant.
Correlations are built on top of $\ket{ \Phi_0 }$ by the action of $e^{ \hat{T} }$, where the cluster operator $\hat{T}$ is expanded as a combination of $n$-particle-$n$-hole ($n$p-$n$h) excitation operators $T_n$.
A simple truncation of the cluster operator, named CC at the singles and doubles level (CCSD), includes terms up to $2p$-$2h$ excitations, \ie{}, $T = T_1 + T_2$.
The $T$ operators read~\cite{Hagen2014Review}
\begin{subequations}
\begin{align}
    & T_1 = \sum_{ai} t^{a}_{i} c_a^{\dagger} c_i\, , \\
    & T_2 = \frac{1}{4} \sum_{abij} t^{ab}_{ij} c_a^{\dagger} c_b^{\dagger} c_j c_i\,  ,
\end{align}
\end{subequations}
where indices $i,j$ and $a,b$ denote single-particle states that are occupied (holes) and unoccupied (particles) in the reference state, respectively.

There is much freedom in choosing the reference state $\ket{ \Phi_0 }$. For closed-shell nuclei a symmetry-conserving reference state is the natural starting point~\cite{Hagen2014Review}, while for open-shell nuclei it is advantageous to break either rotational symmetry~\cite{Novario2020,Hagen2022,Sun2024,Hu2024N50} or particle-number conservation~\cite{Signoracci2015,Tichai2024} when working in a single-reference framework. 
While less important for size-extensive quantities such as binding energies~\cite{papenbrock2024}, the broken symmetries should eventually be restored when computing excited states and their transitions~\cite{Duguet_2015,Duguet_2017,Hagen2022,Sun2024}.

In the Bogoliubov CC (BCC) method, pairing correlations are included self-consistently in the reference state, hence allowing to study ground-state energies of open-shell even-even nuclei~\cite{Signoracci2015,Tichai2024}. 
As pointed out above, this approach breaks particle-number conservation in the reference state.
In BCC, the cluster operator is expressed in the quasi-particle basis, \eg{}
\begin{align}
    \mathcal{T}_2 = 
    \frac{1}{24} \sum_{p_1p_2p_3p_4} t_{p_1p_2p_3p_4} 
    \beta^\dagger_{p_1} \beta^\dagger_{p_2} \beta^\dagger_{p_3} \beta^\dagger_{p_4} \, ,
\end{align}
where the quasi-particle operators $\{\beta^\dagger_k, \beta_k\}$ are obtained from a Bogoliubov transformation that mixes single-particle creation and annihilation operators~\cite{Signoracci2015,Duguet_2017}.
As a consequence particle-hole symmetry is lost, and all indices run over the entire single-particle basis.

\begin{figure}[t!]
    \centering
    \includegraphics[width=1.0\columnwidth]{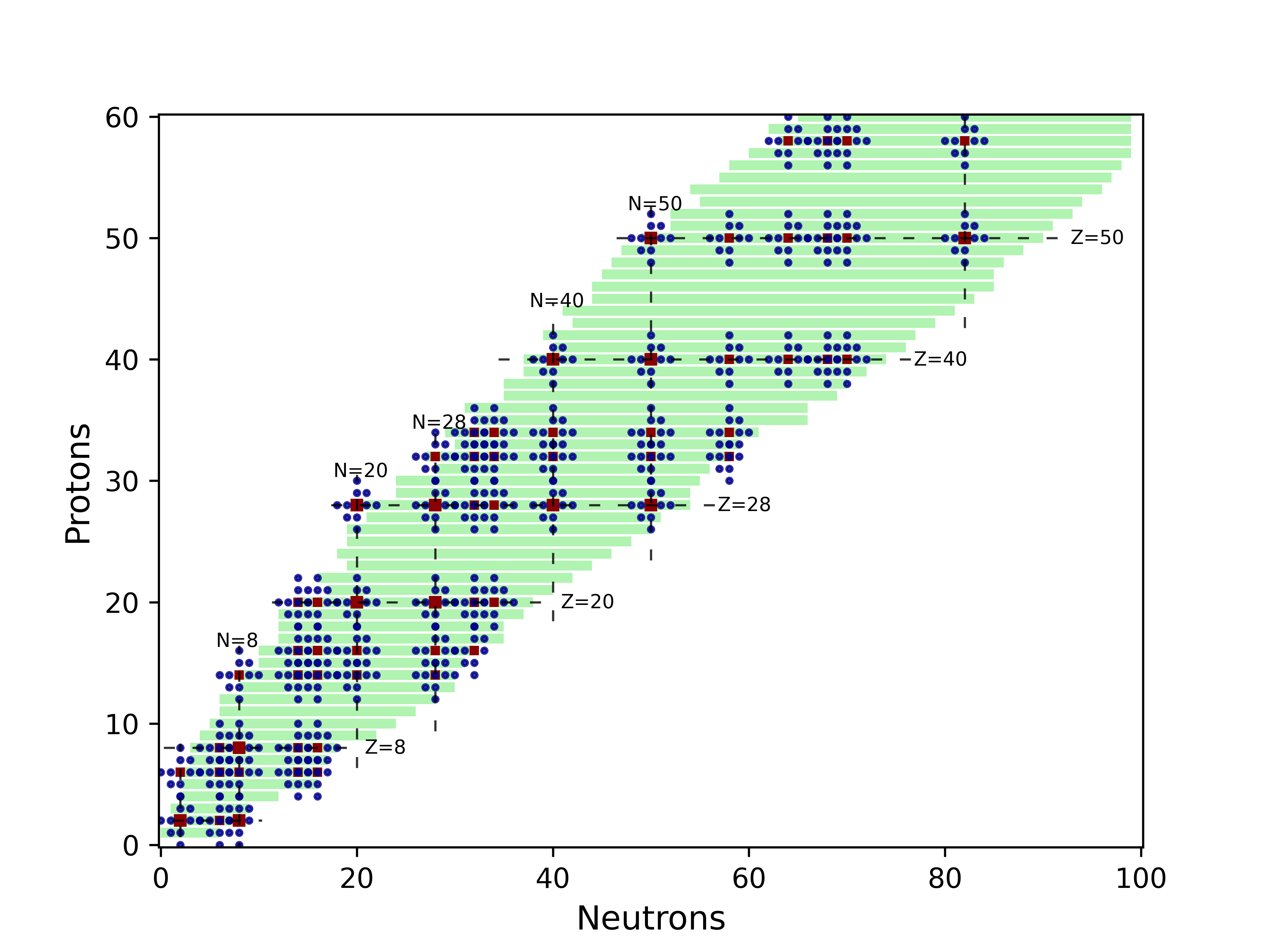}
    \caption{Lower corner of the nuclear chart, with the neutron (proton) number shown in the horizontal (vertical) axis. Closed-shell nuclei are indicated in red, and magic numbers, corresponding to major shell closures, are highlighted with dashed lines. Blue circles denote nuclei that can be reached by using the EOM ansatz. 
    }
    \label{fig: nuclear chart}
\end{figure}

A different approach which does not break any symmetries is rooted in the equations-of-motion (EOM) technique, and provides an efficient and conceptually simple way of studying the spectrum and the response of open-shell systems in the vicinity of a closed-shell nucleus.
In particular, we focus on the two-particle-removed (2PR) ansatz~\cite{Jansen2011}, which allows tackling nuclei that differ by two holes from a shell closure.
In the EOM scheme the states of the $A-2$-nucleus, which satisfy the \Sch equation $\hat{H} \ket{\Psi_f^{(A-2)} } = E_f \ket{\Psi_f^{(A-2)} } $, are interpreted as excited states of the closed-shell neighbour, whose ground state $\ket{ \Psi_0^{(A)} }$ has been determined with a preliminary CC calculation. 
The $ \ket{\Psi_f^{(A-2)} }$ states are assumed to be connected to $\ket{ \Psi_0^{(A)} }$ by the action of an excitation operator $\hat{R}^{(A-2)}_{f}$,
\begin{align}
    \ket{\Psi_f^{(A-2)} }  = \hat{R}^{(A-2)}_{f} \ket{ \Psi_0^{(A)} },
\end{align}
which comprises zero-particle two-holes ($0p$-$2h$) and one-particle three-holes ($1p$-$3h$) contributions,
\begin{align}
    \hat{R}_{f}^{(A-2)} = 
    \frac{1}{2} \sum_{ij} r_{ij} c_j c_i + \frac{1}{6} \sum_{ijka} r_{ijk}^{a} c_a^\dagger c_k c_j c_i.
\end{align}
The amplitudes $ r_{ij}$, $r_{ijk}^{a}$ and energies $E_f$ are determined for the first few low-lying states of the $A-2$ nucleus using the efficient Arnoldi algorithm~\cite{Hagen2014Review}. 
In analogy to the 2PR approach, also the two-particle-attached (2PA)~\cite{Jansen2011,Jansen2013,Bonaiti2024,Brandherm2024} and one-particle-attached/removed~\cite{Gour2006,Hagen2014Review} methods can be constructed. 
By applying the various EOM schemes, a more thorough description of the nuclear chart can be obtained, see Fig.~\ref{fig: nuclear chart}.

Combining the EOM ansatz with the Lorentz integral transform~\cite{Efros1994} technique makes it possible to determine response functions $R(\omega)$ in both closed-~\cite{Bacca2013,Bacca:2014rta,Miorelli:2016qbk,Miorelli2018} and open-shell nuclei~\cite{Bonaiti2024}.
For example, in the case of 2PR systems~\cite{Marino_prep}, the response under the action of an external probe $\hat{\Theta}$, \eg{}, the electric dipole operator, is given by 
\begin{align}
    R(\omega) =
    & \sum_{f} 
    \bra{ \Phi_0^{(A-2)} }  \overline{ \Theta}^\dagger  \ket{ \Phi_f^{(A-2)} }  
    \bra{ \Phi_f^{(A-2)} }  \Bar{\Theta}         \ket{ \Phi_0^{(A-2)} } 
    \nonumber \\
    & \times \, \delta \left( E_f^{(A-2)} - E_0^{(A)} - \omega \right),
\end{align}
where $\Bar{\Theta} \equiv e^{-\hat{T}} \hat{\Theta} e^{\hat{T}} $ and $ \ket{ \Phi_f^{(A-2)} } = \hat{R}^{(A-2)}_{f} \ket{ \Phi_0^{(A)} } $.
In the following, we will focus on the electric dipole polarizability, which can be calculated from the dipole response as the inverse energy-weighted sum rule
\begin{align}
    \alpha_D = 2 \alpha \int d\omega \, \frac{R(\omega)}{\omega}.
\end{align}
The dipole polarizability $\alpha_D$ is subject to extensive theoretical and experimental research due to its connection to nuclear matter properties (\eg{}, the symmetry energy) that are relevant for astrophysical phenomena~\cite{rocamaza2018}.

\section{Numerical results}
\label{sec: results}

\begin{figure}[t]
    \centering
    \includegraphics[width=1.\columnwidth]{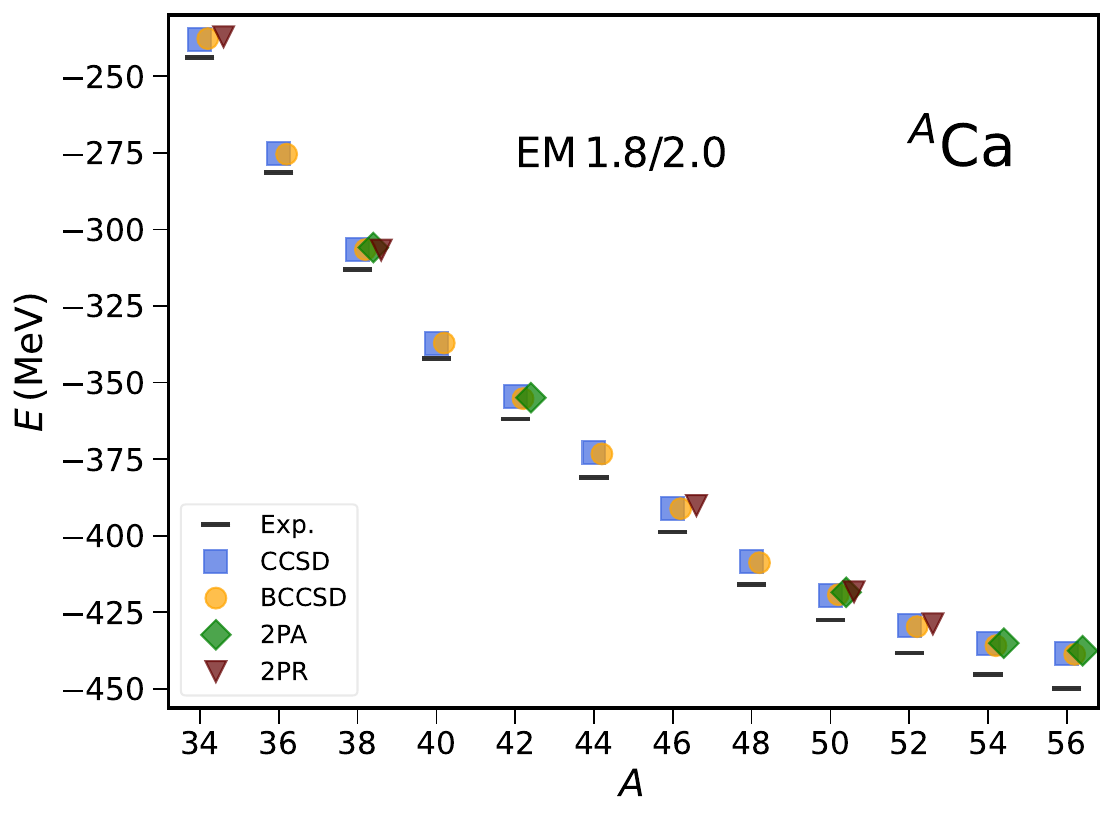}
    \caption{Binding energies of even-mass Ca isotopes as a function of the mass number $A$. 
    The results of the 2PA (diamonds) and 2PR (triangles) CC approaches are compared to Bogoliubov CCSD (BCCSD, circles)~\cite{Tichai2024}, single-reference CCSD
    (squares) calculations~\cite{Novario2020}, and to experimental data from~\cite{wang2021ame}. 
    All computations are performed in a harmonic oscillator basis with $N_{max}=12$ and $\hbar \omega=16$ MeV using the 1.8/2.0 (EM) nuclear interaction~\cite{MagicInteraction}.
    Three-body matrix elements are included up to $E_{max}^{(3)}=16$.
    }
    \label{fig: Ca magic}
\end{figure}

In Fig.~\ref{fig: Ca magic}, we show binding energies of selected even-mass nuclei along the Ca isotopic chain for the 1.8/2.0 (EM) chiral interaction~\cite{MagicInteraction}.
The excellent agreement between CC variants is an important consistency check for the theory.
It suggests that the $1p-3h$ and $3p-1h$ truncations, employed for the 2PR (triangles) and 2PA (diamonds) calculations, respectively, are similar to the singles and doubles level of approximation used in single-reference CCSD (squares) for ground-state energies. Also, comparable results are obtained within Bogoliubov CCSD (BCCSD, circles). It is noteworthy, in particular, that the 2PR ansatz matches well with 2PA-CC in $^{38}\rm{Ca}$ and $^{50}\rm{Ca}$, which are accessible to both techniques.

BCCSD and CCSD are compatible throughout the isotopic chain. We note that these two schemes are equivalent only in the case of closed-shell nuclei. In open-shell systems, both BCCSD and CCSD results lack restoration of broken symmetries. However, restoration of particle number and rotational symmetries 
yield only a small correction to the binding energies~\cite{Hagen2022}. 

While CC built on symmetry-broken states (including BCCSD and CCSD) have a wider reach, the EOM technique provides a similar accuracy at a much smaller computational cost.
Theoretical predictions are overall consistent with experimental data. We observe a small underbinding, which may be slightly corrected by employing larger model spaces. In this regard, we plan to perform 2PA and 2PR calculations including a larger number of three-body states (up to $E_{max}^{(3)}=24$)~\cite{Marino_prep}. 
Ultimately, though, approximate triples corrections~\cite{Hagen2014Review} may be needed to reach an excellent agreement with measurements. 

\begin{figure}[t!]
    \centering
    \includegraphics[width=0.9\columnwidth]{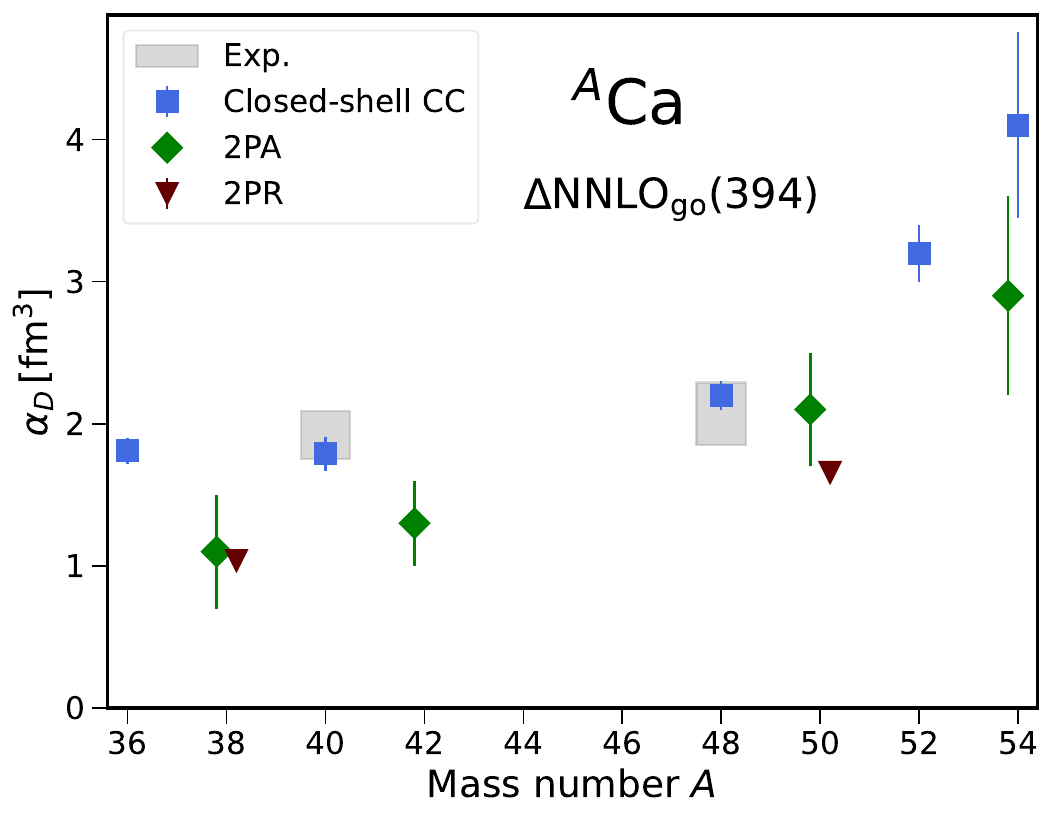}
    \caption{
    Electric dipole polarizability in Ca isotopes. 
    Preliminary calculations performed with 2PR-CC (triangles) are compared to 2PA-CC (diamonds) and closed-shell CC results (squares), and experimental data (grey boxes).
    Error bars accounting for the many-body truncations are reported for 2PA-CC and closed-shell CC. For the case of 2PR, results are shown for $N_{max}=12$ and $\hbar \omega=16$ MeV.
    }
    \label{fig: alphaD}
\end{figure}

The EOM approach can also be employed to tackle dynamical properties of open-shell nuclei.
In Fig.~\ref{fig: alphaD}, we focus on the electric dipole polarizability $\alpha_D$ in the Ca chain.
Preliminary calculations obtained with 2PR-CC (triangles) are compared to 2PA-CC (diamonds) and closed-shell CC (squares) results from~\cite{Bonaiti2024}.
In this case, the $\Delta \rm{NNLO_{\mathrm{GO}}(394)}$ potential~\cite{DeltaGo2020} is employed.
2PA and closed-shell CC include theoretical error bars accounting for the many-body and model-space truncations.
The 2PR results are obtained at a fixed model space ($N_{max}=12$, $\hbar \omega=16$ MeV). More work will be devoted in the future to assess the complete theoretical uncertainty in terms of the many-body and model-space truncations. Even at this level, we can clearly see that the 2PR calculations are compatible with 2PA in $^{38}\rm{Ca}$ and $^{50}\rm{Ca}$, within the already assessed 2PA uncertainties. This benchmark makes us confident about our method, and extensive calculations of the response will soon be performed with 2PA and 2PR.

\section{Conclusions and perspectives}

The equation-of-motion coupled-cluster approach is a flexible method for studying the structure and dynamical response of open-shell nuclei close to magicity.
We have extended the formalism to two-particle-removed nuclei and reported benchmark calculations of binding energies and electric dipole polarizabilities in Ca isotopes. 
Full-scale computations in both even and odd nuclei are ongoing and will allow us to obtain a more comprehensive description of the structure and response of medium-mass nuclei.

\section{Acknowledgements}
This work was supported by the Deutsche Forschungsgemeinschaft (DFG, German Research Foundation) –
Project-ID 279384907 – SFB 1245, through the Cluster of Excellence “Precision Physics, Fundamental Interactions, and Structure of Matter” (PRISMA+ EXC
2118/1, Project ID 39083149), the European Research Council (ERC) under the European Union's Horizon 2020 research and innovation programme (Grant Agreement No.~101020842), by the Research Foundation Flanders (FWO, Belgium, grant 11G5123N) and by the U.S. Department of Energy, Office of Science, Office of Nuclear Physics, under the FRIB Theory Alliance award DE-SC0013617, and Office of Advanced Scientific Computing Research and Office of Nuclear Physics, Scientific Discovery through Advanced Computing (SciDAC) program (SciDAC-5 NUCLEI). This research used resources of the Oak Ridge Leadership Computing Facility located at Oak Ridge National Laboratory, which is supported
by the Office of Science of the Department of Energy
under contract No. DE-AC05-00OR22725. Computer
time was provided by the Innovative and Novel Computational Impact on Theory and Experiment (INCITE)
program, the supercomputer Mogon at Johannes
Gutenberg Universit\"at Mainz and an allocation at the Jülich Supercomputing Center.

\bibliography{bibliography.bib} 

\end{document}